\documentstyle[12pt,epsf]{article}
\def\o  {\omega}

\def\p  {\pi}
\def\P  {\Pi}

\newcommand{\ba}{\begin{array}}
\newcommand{\ea}{\end{array}}
\newcommand{\beq}{\begin{equation}}
\newcommand{\eeq}{\end{equation}}
\newcommand{\bea}{\begin{eqnarray}}
\newcommand{\eea}{\end{eqnarray}}
\newcommand{\beal}{\setcounter{letter}{1} \begin{eqnarray}}
\newcommand{\eeal}{\addtocounter{equation}{1} \end{eqnarray}}

\newcommand{\larrow}{\,\,\,\,\hbox to 30pt{\rightarrowfill}
\,\,\,\,}
\newcommand{\slarrow}{\,\,\,\hbox to 20pt{\rightarrowfill}
\,\,\,}

\newcommand{\IR}{{\rm I\kern-.22em R}}
\begin{document}

\begin{titlepage}
\renewcommand{\thefootnote}{\fnsymbol{footnote}}
\renewcommand{\baselinestretch}{1.3}
\medskip

\begin{center}
{\large {\bf Discrete Spectra of Charged Black Holes}}
\medskip
%\vfill

\vspace{0.5cm}

\renewcommand{\baselinestretch}{1}
{\bf
Andrei Barvinsky $\dagger$
Saurya Das $\flat$
Gabor Kunstatter $\sharp$
\\}
\vspace*{0.50cm}
{\sl
$\dagger$ Theory Department\\
Lebedev Physics Institute\\
Leninsky Prospect 53, Moscow 117924, Russia\\
{[e-mail: barvin@lpi.ru]}\\ [5pt]
}
{\sl $\flat$ Dept. of Mathematics and Statistics \\
University of New Brunswick,  \\
Fredericton, New Brunswick,  Canada E3B 5A3 \\
{[e-mail: saurya@math.unb.ca] } \\ [5pt]
}
{\sl
$\sharp$ Dept. of Physics and Winnipeg Institute for
Theoretical Physics, University of Winnipeg\\
Winnipeg, Manitoba, Canada R3B 2E9\\
{[e-mail: gabor@theory.uwinnipeg.ca]}\\[5pt]
 }

\end{center}

\renewcommand{\baselinestretch}{1}

\begin{center}
{\bf Abstract}
\end{center}
Bekenstein proposed that the 
spectrum of horizon area of quantized 
black holes must be discrete and uniformly spaced.
We examine this proposal in the context of spherically
symmetric charged black holes in a general class of gravity theories.  
By imposing suitable boundary conditions on the reduced phase 
space of the theory to incorporate the thermodynamic properties
of these black holes and then performing a simplifying canonical 
transformation, we are able to quantize the system exactly. 
The resulting spectra of horizon area, as well as that of charge 
are indeed discrete.
Within this quantization scheme, near-extremal black holes (of any mass)
turn out to be highly quantum objects, whereas extremal black holes
do not appear in the spectrum, a result that is consistent with the 
postulated third law of black hole thermodynamics.


\end{titlepage}
\clearpage
\section{Introduction}

Black holes serve as interesting theoretical laboratories for testing
the validity and predictive power of theories of quantum gravity.
The robustness of thermodynamic laws associated with black
holes, along with their inherent structural simplicity require that any
reasonable theory ought to  predict some of their generic features. Notable
among them are Bekenstein-Hawking entropy associated with horizon
area of black holes, and Hawking radiation from those horizons \cite{BH}.
Together, they ensure that a generalized second of thermodynamics is
valid in the presence of black holes.
This law was proposed by Bekenstein in the early seventies \cite{second}.

Along with these spectacular observations, there also arose the question
as to whether the spectra of observables related to black holes were continuous
or discrete. Again, by a remarkable set of thought experiments, it was inferred
by Bekenstein and collaborators that provided a black hole is far away
from extremality, its horizon area can be regarded as an adiabatic invariant.
Now, it is well known from quantum theory, that adiabatic
invariants are always quantized, and their spectra are equally spaced
\cite{pauling}. Naturally, they
proposed that horizon area spectrum is discrete and of the form
\cite{bm0} :
\bea
a_n =n a_0 ~~,
\label{bs1}
\eea
where $a_0$ is a fundamental quantum of area.  Recently, Bekenstein and
collaborators
did an algebraic analysis, and showed that under certain plausible assumptions,
the area spectrum was indeed discrete and equally spaced \cite{bm}.
This feature
was confirmed by several other analyses as well, which leads one to believe
that spectra of this kind should be features of all black holes.
In fact, discrete spectra such as above was found by many authors using
diverse approaches \cite{kogan,berezin,rovellismolin,kastrup,louko,bk,VW,
garattini,kastrup_strobl,poly,MRLP,bdk1}.

In this article, we try to address these questions from a slightly 
different perspective. 
In particular, we consider spherically symmetric charged black holes in a 
generic theory of gravity. The reduced (physical) phase 
space is just four dimensional. We Euclideanize and impose periodic 
boundary conditions that reflect the thermodynamic nature of the black holes. 
By making a canonical transformation, 
we are able to quantize the 
resulting theory {\it exactly} and obtain a spectrum similar to that 
proposed by Bekenstein (\ref{bs1}) for uncharged black holes. 
Moreover, our spectrum predicts  a remnant ground state,
so that any physical process such as Hawking radiation must stop when this 
minimum value of horizon area is reached.

The article is arranged as follows: in the next section, we elaborate
on the exact quantization procedure adopted by us
(following an analysis developed in \cite{bk}), and the spectrum obtained
thereof. In section (\ref{adiabaticinvariant}), we present a rigorous
derivation of a slightly modified version of Bekenstein's hypothesis
that the horizon area is an adiabatic invariant. Then,
in section (\ref{secdry}), we go one step further and show
why our spectrum
ought to agree with Bekenstein's one, at least in the uncharged case, by
formulating an exact correspondence between the operators in his algebra
and the fundamental gravitational degrees of freedom in our analysis.
Finally, in section (\ref{concl}) we examine some consequences of our
spectrum and conclude by stating some open problems. For further details
and intermediate steps, we refer the reader to the original papers
\cite{bdk1,dry}.

\section{Reduced Phase Space Quantization}

We start with a generic theory of gravity which describes the dynamics
of charged black holes (e.g. it could be Einstein-Maxwell theory, or those
that arise in low energy string theory). Since we are primarily
interested in the
generic and robust features of these black holes, we rid ourselves
of unnecessary complications, and consider only those solutions which are
spherically symmetric. 
It is by now well known that under the above assumptions, 
all such gravity actions
can be dimensionally reduced to a simple particle mechanics action in 
$1$-dimension of the form \cite{dk,mk,shelemy}:
\bea
    I^{red}= \int dt \left(P_M \dot{M} +
    P_Q \dot{Q} - H(M,Q)\right) \, ,      \label{reduced action}
\eea
where $M$ and $Q$ are the mass and charge of the black hole respectively, and
$P_M$ and $P_Q$ are the conjugate momenta. The boundary conditions used
were those of \cite{lw}.
The above action automatically ensures that $M$ and $Q$ are constants of motion,
as required by the generalized Birkhoff theorem. It can be shown that
the momentum $P_M$ has the interpretation of asymptotic Schwarzschild
time difference between the left and right wedges of a Kruskal diagram
\cite{kuchar,thiemann,gkl}.
The reduced phase space of spherically symmetric solutions of generic 
gravity-electromagnetism systems is thus four dimensional, spanned 
by the coordinates $(M,Q,P_M,P_Q)$. Furthermore, we restrict the mass parameter
to be non-negative. Note that this describes not only black hole geometries,
but also other objects such as spherical stars. Since we are 
interested in the quantum mechanics of black holes, we restrict the phase
space to that of 
the latter by making use of the following fact: $P_M$, the conjugate 
to the mass variable, is effectively the 
asymptotic ``Schwarzschild'' time, and is periodic in the
Euclideanized formulation of black hole
thermodynamics (with period equal to the inverse Hawking temperature $T_H$
times $\hbar$). We therefore go to the Euclidean sector of the 
theory and impose the additional restriction that $P_M$
is periodic with the same period. That is:
\bea
P_M \sim P_M + \frac{\hbar}{T_H(M,Q)}~~.
\eea
Although this may seem ad-hoc at this point, we will see that this one
simple (and plausible) assumption helps in deriving satisfactory spectra
for both area and
charge of the black hole. Also, similar assumptions regarding periodicity
(or associating a fundamental time scale with black holes)
were made using somewhat complex arguments in the past
\cite{kastrup,louko,kastrup_strobl}.
Here, we simply proceed
with this assumption, and note that this restricts the subspace $(M,P_M)$
into a wedge like region, bounded by the $M$ axis and the locus of points
$P_M=\hbar/T(M,Q)$ with varying $M$ (e.g. for a Schwarzschild black hole
in four dimensions,
it is the straight line $P_M=8\pi G_4 M, G_4$ being four
dimensional Newton's constant).
The appearance of a wedge
may seem a little disturbing, but it was shown in \cite{bk,bdk1} that this
wedge can be removed by making the following canonical transformation on
the full phase-space:
\bea
    &&X=\sqrt{\hbar (S_{BH}(M,Q)- S_0(Q))\over \pi}
    \cos(2\pi P_M T_H(M,Q)/\hbar) \, , \label{ctr1} \\
    &&\Pi_X= \sqrt{\hbar (S_{BH}(M,Q)- S_0(Q))\over \pi }
    \sin (2\pi P_M T_H(M,Q)/\hbar)    \, ,        \label{ctr2}\\
    &&Q=Q \, ,  \\
    &&\P_Q=P_Q + \Phi P_M + S_0'(Q) P_M T_H\, ,                 \label{ctr4}
\eea
where $S_{BH}(M,Q)$ is the Bekenstein-Hawking entropy of the black hole and
$' \equiv d/dQ$.
$S_0(Q)$ is the value of $S$ attained at extremality as the mass of the
black hole approaches its charge. For example, for Reissner-Nordstr\"om
black holes in $d$ spacetime dimensions,
 \bea
    S_0(Q)= K_{(d)}Q^{(d-2)/(d-3)}/\hbar \, , \label{G(Q)}
\eea
where
    \bea K_{(d)} = (1/4) (A_{d-2}/G_{d})^{(d-4)/2(d-3)}
    (8\pi/(d-2)(d-3))^{(d-2)/2(d-3)} \, ,\label{Kd} \eea
($A_{d-2}=2\pi^{(d-1)/2}/\Gamma((d-1)/2))$ is the area of the
unit $d-2$ sphere). It is interesting to note that in all cases
except $d=4$, the entropy bound depends explicitly on the
gravitational constant $G_d$.
$S_0(Q)$ appears in the transformation in order to guarantee that the 
square-root remains real for all values 
of the parameters $M$ and $Q$ that correspond
to physical black holes (as opposed to naked singularities). 
Squaring and adding (\ref{ctr1}) and (\ref{ctr2}), we get:
\bea
S_{BH}-S_0(Q) = {2 \pi\over \hbar}
    \left({X^2\over2} +
    {\P_X^2\over 2}\right). \label{harmonic oscillator 1}
\eea
The right hand side is immediately recognizable as the
Hamiltonian of a linear harmonic oscillator on the
$(X,\Pi_X)$ subspace.
Quantization is straightforward (with usual
identifications $ {\hat X} \rightarrow X~,~
{\hat \Pi}_X \rightarrow -i \partial/\partial X$), yielding:
\bea
S_{BH} = 2\pi \left(n + \frac{1}{2} \right) + S_0(Q) ~~~,~n=0,1,2, \dots  \label{quant1}
\eea
Before proceeding further, we note that the above spectrum
automatically satisfies the extremality bound $S_{BH} \geq
S_0(Q)$. In fact, with our choice of factor ordering, it is a
strict inequality. Although one might argue that this classical
bound may be modified (or even violated) for microscopic black
holes, for large black holes, it should evidently hold. Our
spectrum ensures that this is indeed the case. Also, note that the
(\ref{quant1}) implies that (near)-extremal black holes are highly
quantum mechanical objects, irrespective of their mass, since they
correspond to low values of the quantum number $n$. Although it is
somewhat counter-intuitive to think of large, near-extremal black
black holes as quantum mechanical, this view is consistent with
the thermodynamic interpretation of black holes, since
(near)-extremal black holes are associated with extremely small
temperatures, signifying transition to quantum regimes. The
quantum nature of black holes near extremality, and the breakdown
of macroscopic laws in this regime were also found earlier
\cite{ddr,dmb}.

Another potentially important feature of the spectrum (\ref{quant1})
is that, with our choice of factor ordering, 
extremal black holes ($S_{BH}= S_0(Q)$) are not in the quantum spectrum. 
This suggests that
it may not be possible for non-extremal black holes to decay, or even get 
arbitrarily close, to extremality. This intriguing possibility has been
discussed recently by Medved in \cite{medved} who used duality arguments to
conclude that  back reaction effects prevent Reissner-Nordstrom type black holes in any dimensions from reaching the extremal state.

To complete the analysis of the spectrum, we use the
following result from \cite{dk}:
$$ \delta P_Q = -\Phi P_M + \delta \lambda~~,$$
where $\Phi$ is the electrostatic potential on the boundary under
consideration, and variation refers to small change in boundary conditions,
$\lambda$ being the gauge parameter at the boundary.
This in turn implies that for compact $U(1)$ gauge group,
$ \chi \equiv e\lambda/\hbar = e (P_Q + \Phi P_M)$ is periodic
with period $2\pi$ ($e=$ electronic charge).
Also, we saw earlier from thermodynamic arguments
that $\alpha \equiv 2\pi P_M T_H(M,Q)$
has period $2\pi$. In terms of these
`angular' coordinates, the momentum $\Pi_Q$ in (\ref{ctr4}) can be written as:
$$ \Pi_Q = \frac{\hbar}{e} \chi + \frac{\hbar}{2\pi} S_0'(Q) \alpha~~.$$
Thus, the following identification must hold in the $(Q,\Pi_Q)$ subspace:
\bea
\left(Q, \Pi_Q \right) \sim
\left( Q, \Pi_Q + 2\pi n_1 \frac{\hbar}{e} + n_2 \hbar S_0'(Q) \right)~~,
\label{qpq}
\eea
for any two integers $n_1,n_2$. Now, wavefunctions of charge eigenstates
are of the form:
$$ \psi_Q(\Pi_Q) = \exp\left(i Q\Pi_Q/\hbar \right) ~~,$$
which is single valued under the identification (\ref{qpq}),
provided there exists another integer $n_3$ such that:
$$ n_1 \frac{Q}{e} + n_2 \frac{Q}{2\pi} S_0'(Q) = n_3~~.$$
Now, it can be easily shown  that
the above conditions is satisfied if and only if the following
two quantization conditions hold:
\bea
\frac{Q}{e} &=& m \label{chq1} \\
\frac{Q}{2\pi} S_0'(Q) &=& p ~~, \label{chq2}
\eea
where $m$ and $p$ are any two integers.
While the first condition is the familiar
charge quantization condition, the second is a new constraint
on the $U(1)$ charge. For example, for Reissner-Nordstr\"om
black holes, Eq.(\ref{G(Q)}) implies:
\bea
\frac{K_{(d)}}{2\pi} \left( \frac{d-2}{d-3} \right)
\frac{Q^{(d-2)/(d-3)}}{\hbar} = p . \label{quant3}
\eea
Together (\ref{quant1}) and (\ref{quant3}) imply that the horizon area
spectrum of the Reissner-Nordstr\"om black hole is given by:
\bea
S_{BH} = 2\pi \left[ n + \left(\frac{d-3}{d-2}\right) p \right]
+ \pi ~~. \label{npspectrum}
\eea
Using the Bekenstein-Hawking entropy formula
\bea
S_{BH} = \frac{A_{BH}}{4~\ell_{Pl}^{d-2}} ~
\eea
(where $\ell_{Pl}$ is the Planck length in $d$-dimensions),
(\ref{npspectrum}) gives the following spectrum of its horizon area
\bea
A_{BH} =
8\pi \left[ n + \left(\frac{d-3}{d-2}\right) p \right]~\ell_{Pl}^{d-2}
+ 4\pi~\ell_{Pl}^{d-2} ~~.
\label{npspectrum1}
\eea
This is our main result. The quantum number $p$ determines the charge of
the quantum black hole, while $n$ determines the excitation of the black hole
above extremality. As mentioned earlier, 
although classically the extremality bound can be
reached, our analysis predicts the remarkable feature that this classical
bound is never saturated due to small vacuum fluctuations of the horizon.
Also, note that the ground state of the spectrum is at:
\bea
A_{BH} (n=0=p) = 4 \pi \ell_{Pl}^{d-2} ~~,
\eea
implying that there is a `zero-point area' of Planckian dimensions.
This also implies that if for example Hawking evaporation radiates away
the energy (and area) of a black hole, then it must stop at the above
value. It is tempting to speculate that this Planck sized remnant will
retain information that fell into the black hole earlier, thus avoiding the
information loss `paradox'. Such remnants have been
anticipated in many early works in quantum gravity and astrophysics, and
there remains a lively debate about their existence in general
\cite{remnants}.

\section{Adiabatic Invariants}
\label{adiabaticinvariant}

Having derived the spectrum (\ref{npspectrum1}), we return to Bekenstein's
original reasoning about discrete area spectrum from adiabatic invariants.
From a class of novel thought experiments he argued that horizon areas of
black holes with charge must be adiabatic invariants. However, here a very
similar result can be derived from first principles: consider
Eq.(\ref{harmonic oscillator 1}). Since the right hand side
describes a harmonic oscillator, the
periodic orbits in phase space naturally give rise to the following adiabatic
invariant:
\bea
{\cal J} = \oint \Pi_X dX = \frac{A- 4G \hbar S_0(Q) }{4G/\pi}~~.
\label{refine}
\eea
Thus for $S_{BH} \gg S_0$ (i.e. far from extremality), the horizon area is
indeed an adiabatic invariant. However, close to extremality, the above
relation suggests that it is the area above extremality which is an
adiabatic invariant. We interpret this as a slight refinement
over Bekenstein's original hypothesis. The advantage of relation
(\ref{refine}) is that on the one hand it is consistent 
with the discrete spectra
(\ref{npspectrum1}), and on the other hand, it ensures
that the extremality bound $S_{BH} \geq S_0$ is always obeyed.

\section{Relation to Bekenstein's Analysis}
\label{secdry}

Now, we examine the relation of our spectrum to that derived by Bekenstein
from an algebraic point of view in \cite{bm}. The
issue was examined in \cite{dry} for uncharged black holes, and we review the
results here. The extension to charged black holes will be left to a future
publication. In \cite{bm}, Bekenstein and collaborators
proposed the existence of a set of linear
operators $\{ {\hat A}, {\hat {\cal R}}_{ns_n} \}$, where the first operator
corresponds to the horizon area observable, and the second creates a single
black hole state from vacuum with area $a_n$,
in an internal quantum state $s_n$.
It is assumed that $ s_n \in \{0,1, \dots, e^{a_n} - 1 \}$, to account
for the internal degeneracy associated with the Bekenstein-Hawking entropy.
Symmetry, linearity and closure imply that the
algebra between these fundamental operators must be of the form:
\bea
\left[ {\hat A}, {\hat {\cal R}}_{n s_n} \right] &=&
a_n {\hat {\cal R}}_{n s_n} ~~, \label{dry3} \\
\left[ {\hat A}, {\hat {\cal R}}^\dagger_{n s_n} \right] &=&
- a_n {\hat {\cal R}}^\dagger_{n s_n} ~~, \label{dry4} \\
\left[  {\hat A},
\left[   {\hat {\cal R}}^\dagger_{m s_m},
{\hat {\cal R}}_{n s_n}  \right]
\right] &=&
(a_n - a_m) \left[ {\hat {\cal R}}^\dagger_{m s_m},
{\hat {\cal R}}_{n s_n}  \right]  ~~~ \mbox{iff}~a_n > a_m~~, \label{dry5} \\
\left[ {\hat {\cal R}}_{n s_n}, {\hat {\cal R}}_{m s_m} \right]
&=& \epsilon^k_{nm}
{\hat {\cal R}}_{k s_k}
~~(\epsilon^k_{nm} \neq 0, ~~\mbox{iff}~a_n + a_m = a_k ) ~~. \label{dry6}
\eea
Now, it was shown in \cite{bm} that the spectrum of the above algebra
involves both addition and subtraction of area levels, which is possible
if and only if the area levels  are equally spaced; i.e.,
\bea
a_n = n {a_0} + {\bar a} ~~~~~~~n=0, 1, 2, \cdots~~
\eea
where ${\bar a}$ is a constant which can
take any arbitrary value. In \cite{bm}, ${\bar a}$
was set to zero so that the ground state area. 
However, the algebra Eqs.(\ref{dry3}-\ref{dry6}) does not in any way impose
such a constraint, and in fact remains unchanged even if there is a
non-zero ${\bar a}$. For example, for
Reissner-Nordstrom black holes, $a_0$ and ${\bar a}$ were chosen to be
\cite{dry} :
\bea
a_0 = 2 {\bar a} = 8 \pi \ell_{Pl}^{d-2} ~~.
\eea
With this identification,
%(and with the entropy formula
%$S_{BH} = A/4G_d\hbar$), 
the above spectrum becomes identical to the uncharged
version of (\ref{npspectrum1}). Although the operators in
Eqs.(\ref{dry3}-\ref{dry6}) have so far been kept abstract, the
picture can be completed by their explicit construction:
\bea
{\hat {\cal R}}_{ns_n} &=& (P^\dagger)^n~{\hat g}_{s_n}~~, \\
{\hat A} &=& ({\hat P}^ \dagger {\hat P} + 1/2){\bar a}~~, \\
{\hat P}^\dagger &=& \frac{1}{\sqrt{2} } \left[ {\hat X} - i {\hat \P_X}  \right]~~,
\eea
which automatically satisfy the algebra:
\bea
[ {\hat P}, {\hat P}^\dagger] &=&1~, \label {pr1}\\
{[}{\hat P}, {\hat g}_{s_m}]&=&[{\hat P}^\dagger, {\hat g}_{s_m}]=0~,
\label{pr2}\\
{[}{\hat g}_{s_m},{\hat g}_{s_n}] &=& \epsilon_{mn}^k {\hat g}_{s_k}
~~~~{\rm where}~ \epsilon_{mn}^k \neq 0 ~{\rm iff}
~ s_k = s_m+s_n~.
\eea
Thus we can see that the operators used by Bekenstein to derive
equally spaced spectra for black hole can be constructed out of
fundamental gravitational degrees of freedom, at least in the context
of spherically symmetric uncharged black holes. This makes our analysis
and results perfectly consistent with those of Bekenstein.

\section{Summary and Conclusions}
\label{concl}

Finally, let us consider the implications of our results in
a physical process in which the black hole emits
a photon by making a quantum jump from one level to the next lower
level.
Assuming that the black hole decays
by emitting just one photon with the lowest allowed frequency
$\o_0$ (for simplicity we assume
uncharged particle emission and four dimensions),
its initial and final masses are
$M + \hbar \o_0$ and $M$ respectively, and using (\ref{npspectrum1}) the
following relation holds :
$$ A_{BH}(M+ \hbar \o_0,Q) - A_{BH}(M,Q) 
= A_{BH}(n+1) - A_{BH}(n) = 4\pi \ell_{Pl}^2 ~~.$$
Using $A_{BH} = 4 \pi r_+^2$ and $r_\pm = G_4 M \pm
\sqrt{(G_4M)^2 - G_4Q^2}$, we get:
\bea
\o_0 = \frac{ (r_+ - r_-) \p}{A}\, .
\label{freq}
\eea
In the $Q\rightarrow 0$ limit, the above frequency agrees
with that found in \cite{bm} up to factors of order unity.
However, it differs significantly from predictions of loop
quantum gravity \cite{rovelli}.

To summarize, in this article we have shown how the spectra of
black hole observables postulated by Bekenstein can be derived within
an explicit, rigorous (given one basic assumption) quantization scheme. Our
derivation pertains to the spherically symmetric charged black hole
sector of generic theories of gravity. Once non-spherically symmetric modes
are introduced, one naturally expects some modifications to the spectrum.
In particular, the quantum numbers that we derived are 
analogous to the principle
quantum numbers of the hydrogen atom. 
Introducing additional modes will bring in 
more quantum numbers, such as angular momentum, and many 
more states which, to a first approximation, are highly degenerate.  
An important open question is to what extent higher order corrections 
break this degeneracy and cause the
spaces between the discrete spectrum values to be 
``filled in'', potentially restoring the continuous radiation 
spectrum of Hawking's original work. 

Another intriguing point to note is that quantization conditions
(\ref{chq1}) and (\ref{quant3}) imply that the fundamental
charge $e$ is constrained by the following relation:
\bea
e^{(d-2)/(d-3)} = \left[ \frac{2\pi \hbar}{K_{(d)}}
\left( \frac{d-2}{d-3}\right) \right]~ \frac{p}{m^{(d-2)/(d-3)}}~~,
\label{fine2}
\eea
which is the $d$-dimensional fine-structure constant. The above relation
can be interpreted in one of the following ways, depending on whether one
considers the black hole or the electronic charge as fundamental
physical entities. According to the first point of view,
even if a single black hole is present in the universe, it would imply
that the electronic charge would have to satisfy the condition
(\ref{fine2}) for some integral values of $p$ and $m$. This is
reminiscent of Dirac's quantization condition, according to which,
presence of a single magnetic monopole in the universe of strength
$g$ would require all electric charges to be quantized in units of
$2\pi\hbar/g$ \cite{dirac}. This is also in the spirit of
Coleman's `Big-Fix Mechanism' in which he argued that wormholes
can fix the constants of nature \cite{bigfix}. Indeed for $d=4$,
condition (\ref{fine2}) translates to
$$ \frac{e^2}{\hbar} = \frac{p}{m^2}~~,$$
which says that the fine structure constant in our universe should
somehow be approximated by the above form.

Alternatively, one can take the viewpoint that electronic charges
are fundamental, so that charged black holes must be created such that
their charges satisfy the conditions (\ref{chq1}) and (\ref{chq2}).
Whichever interpretation one chooses to embrace, 
we have shown that quantum gravity in general, and
black holes in particular, may play a very important role in 
determining
the fundamental constants of nature. 
In any case, it is certainly clear that Professor 
Bekenstein's contributions to the field of black hole
dynamics and to our understanding of the universe in general can
hardly be over-emphasized.

\vspace{1cm}
\noindent {\bf Acknowledgments}

\noindent  S.D. and G.K. would like to thank Viqar Husain for
helpful comments and encouragement. S.D. thanks P. Ramadevi and U.
A. Yajnik for useful correspondence, and collaboration on which
paper \cite{dry} was based. G.K. thanks Valeri Frolov and the
gravity group at the Theoretical Physics Institute, University of
Alberta for useful discussions. We also acknowledge the partial
support of the Natural Sciences and Engineering Research Council
of Canada. This work was also supported by the Russian Foundation
for Basic Research under the grant No 02-01-00930.

\end{document}